\documentclass[pdflatex,sn-mathphys-num]{sn-jnl}

\usepackage{graphicx}%
\usepackage{multirow}%
\usepackage{amsmath,amssymb,amsfonts}%
\usepackage{amsthm}%
\usepackage{mathrsfs}%
\usepackage[title]{appendix}%
\usepackage{xcolor}%
\usepackage{textcomp}%
\usepackage{manyfoot}%
\usepackage{booktabs}%
\usepackage{algorithm}%
\usepackage{algorithmicx}%
\usepackage{algpseudocode}%
\usepackage{listings}%

\theoremstyle{thmstyleone}%
\theoremstyle{thmstyletwo}%

\theoremstyle{thmstylethree}%

\begin{document}

\title[Article Title]{Integrated tunable mid-infrared electro-optic frequency comb generator based on nonlinear conversion}


\author*[1]{\fnm{Pierre} \sur{Didier}}\email{pdidier@phys.ethz.ch}

\author[1]{\fnm{Prakhar} \sur{Jain}}

\author[1]{\fnm{Tristan} \sur{Kuttner}}

\author[1]{\fnm{Oliver} \sur{Pitz}}

\author[1]{\fnm{Rachel} \sur{Grange}}

\affil[1]{ETH Zurich, Department of Physics, Institute for Quantum Electronics, Optical Nanomaterial Group, Zurich, Switzerland}



\abstract{Mid-infrared frequency combs enable highly selective and sensitive molecular spectroscopy by leveraging the strong vibrational transitions in this spectral region. Among these, there is a particular need for compact, tunable sources with electronic control over comb parameters for integrated sensing platforms. In this work, we demonstrate a mid-infrared electro-optic frequency comb source based on nonlinear frequency conversion in thin film lithium niobate. The system combines a near-infrared pump, amplitude-modulated using an integrated Mach–Zehnder modulator for lock-in detection, with a telecom-band electro-optic comb generated via a double-pass phase modulation scheme. Mid-infrared comb generation is achieved through difference frequency generation in a periodically poled waveguide. By tuning the telecom seed laser and the chip temperature, we obtain mid-infrared combs with a bandwidth of approximately 6 nm and center wavelength tunability of over 200 nm. The comb free spectral range is directly controlled via the applied radio-frequency modulation. Operation across multiple integrated photonic circuits reaching wavelengths up to 3.7 \textmu m is demonstrated. Furthermore, dual-tone EO comb generation in the mid-infrared is realized. To our knowledge, this is the first integrated mid-infrared electro-optic comb source offering independent electronic control of both center wavelength and comb spacing.}


\maketitle
\section{Main}\label{sec1}

The mid-infrared (MIR) spectral range, spanning approximately 3-14~\textmu m, offers major benefits for free-space optical applications, including high atmospheric transparency, reduced scattering from micrometer-scale aerosols, and improved resilience to turbulence~\cite{Jony2019,Spitz2021,corrigan2009quantum,liu2019mid,Esmail2017}. In addition, many environmentally and biologically relevant molecules exhibit strong and distinctive vibrational absorption lines in this range, often significantly stronger than in the near infrared (NIR), allowing highly sensitive and selective spectroscopic detection~\cite{haas2016advances,zhang2014applications}. However, there remains a need for highly efficient integrated platforms in order to fully exploit the unique advantages of this spectral range.

Within this broader context, optical frequency combs have become indispensable tools for precision spectroscopy and metrology. Their broadband, evenly spaced, and phase-coherent spectral lines act as optical frequency rulers, enabling dual-comb spectroscopy~\cite{coddington2016dual}, high-resolution metrology~\cite{udem2002optical}, and rapid, parallel detection of multiple absorption features with high sensitivity and spectral resolution~\cite{schilt2003wavelength,picque2019frequency}. A wide range of approaches to frequency-comb generation has been developed, most prominently in the visible and NIR spectral regions~\cite{chang2022integrated}. Among these, mode-locked lasers represent a mature and well-established platform reaching industry implementation~\cite{diddams2007molecular}. In parallel, Kerr nonlinear microresonators have emerged as a powerful alternative, capable of generating octave-spanning spectra and achieving outstanding performance in the NIR~\cite{wang2019monolithic,pfeiffer2017octave,del2007optical} but suffering from instability. Electro-optic (EO) comb generators provide another approach, in which a continuous-wave laser is modulated using RF-driven electro-optic phase modulator. This technique enables combs with excellent stability and tunable repetition rates. To date, many integrated EO comb demonstrations have relied on lithium niobate (LiNbO\textsubscript{3}, LN), owing to its low propagation loss, broad transparency window extending beyond $4.5~\mu\mathrm{m}$, and strong second-order nonlinearity $\chi^{(2)}$, with a Pockels coefficient of up to $33~\mathrm{pm/V}$~\cite{boyd2008nonlinear}. However, conventional bulk LN waveguides fabricated by titanium in-diffusion suffer from weak optical confinement and large bend radii, which limit scalability and device efficiency~\cite{xie2022linbo3}. Thin-film lithium niobate on insulator (LNOI) overcomes these limitations by enabling strong confinement and compact device architectures. As a result, LNOI has driven rapid progress in electro-optic photonic devices, with demonstrated modulation bandwidths exceeding $100~\mathrm{GHz}$ and with a modulation efficiency below \(1~\text{V}\cdot\text{cm}\)~\cite{wang2018integrated,xu2020high}. At 1.5~\textmu m, non-resonant LNOI modulators have achieved EO comb bandwidths of $20~\mathrm{nm}$ with $25~\mathrm{GHz}$ line spacing~\cite{zhang2023power}. In contrast, resonant ring-based devices have demonstrated significantly broader bandwidths of up to $132~\mathrm{nm}$ with $30.925~\mathrm{GHz}$ spacing and conversion efficiencies approaching $30\%$~\cite{rueda2019resonant,hu2022high,zhang2019broadband}. More recently, resonant EO comb generation has also been demonstrated in alternative material platforms like lithium tantalate~\cite{zhang2025ultrabroadband}. While the achievable bandwidth of EO combs is generally limited by modulator efficiency and available RF drive power, they still provide a highly coherent and scalable route that is particularly attractive for integrated photonics. Nevertheless, despite rapid progress in integrated MIR modulators~\cite{didier2026thin,lee2026suspended,montesinos2022mid}, translating electro‑optic comb generation concepts to the MIR remains challenging. This is due to the limited availability of efficient phase modulation at these wavelengths, typically associated with large \(V_{\pi}\cdot L\), due to increased propagation losses, weaker optical confinement, and the relatively low maturity of integrated MIR platform technologies~\cite{guo2018mid}. In particular, efficient integrated EO comb generation in the MIR critically depends on achieving low \(V_{\pi}\cdot L\) \cite{didier2026thin}. 

Quantum cascade laser (QCL) combs are attractive owing to large bandwidth and high power per line, but they are often constrained by noise and stringent thermal management requirements. Several studies have demonstrated MIR comb generation through phase modulation using QCLs~\cite{hugi2012mid}, including quantum-walk QCL combs that rely purely on phase modulation~\cite{heckelmann2023quantum}. More recently, racetrack QCLs have been shown to generate soliton-based pulses with picosecond-scale durations~\cite{kazakov2025driven}, offering a compact route to MIR pulse generation. However, QCL emission is often limited to wavelengths above 3.9 \textmu m, leaving part of the MIR spectral region inaccessible to QCL-based combs. In addition, QCL devices generally exhibit high electrical power consumption and demand stringent thermal management, which further complicates their use in compact sensing platforms. Other approaches include interband cascade laser frequency combs \cite{Schwarz2019,bagheri2018passively,abajyan2025mid}, but they still typically exhibit limited output power, restricted bandwidth, modest tunability of the comb center wavelength, and comparatively high noise levels. Differently, nonlinear frequency conversion from the NIR using bulk periodically poled lithium niobate (PPLN) crystals, although capable of broad spectral coverage, generally relies on bulky free-space setups and requires very high pump powers~\cite{hoghooghi2022broadband}. Some demonstrations have employed an optical parametric oscillator to generate a frequency comb extending up to \(3.3~\mu\mathrm{m}\). But, it requires a high-power pulsed pump source with a fixed free spectral range (FSR) matched to that of the ring resonator.~\cite{roy2023visible}. These limitations motivate the development of alternative MIR comb architectures that combine high coherence, wide tunability, and scalable integration. EO combs are particularly powerful in this regard because their coherence is directly inherited from the seed laser and RF drive, in contrast to QCL‑ and OPO‑based combs, where coherence arises from complex nonlinear dynamics and is therefore more difficult to precisely control and stabilize. In addition, EO combs offer deterministic and independent control over the comb spacing, providing a degree of flexibility that is difficult to access with conventional MIR comb platforms. 

In this work, we demonstrate efficient MIR electro‑optic frequency‑comb generation on an integrated LNOI platform by first generating a NIR EO comb through phase modulation and subsequently converting it to the MIR via difference‑frequency generation in an integrated PPLN section. The resulting MIR comb is tunable over more than 200 nm while maintaining an instantaneous bandwidth of approximately 6 nm around the center wavelength and a free spectral range up to 12.5 GHz, with typical per‑line powers on the order of –52 dBm at the output. In this architecture, the comb spacing is defined by the RF drive frequency, whereas the carrier offset is independently controlled by tuning the signal source and chip temperature, thereby enabling independent control of the MIR center frequency and free spectral range within a single device. Across several devices, we demonstrate MIR comb generation from 3.15 \textmu m up to 3.7 \textmu m. Moreover, we demonstrate operation of our device under dual‑tone RF driving, highlighting its flexibility. This shows that our device can generate an on‑chip dual comb using two independent phase modulators, allows low‑frequency beat notes to be detected with a low‑bandwidth photodetector, and establishes a compact route to MIR spectroscopy without a conventional optical spectrometer. Finally, in this architecture both the telecom‑wavelength comb and the MIR comb are simultaneously available at the output, enabling spectroscopic applications that exploit molecules with shared absorption bands in both the NIR and MIR for enhanced precision.

\begin{figure}[ht!]
    \centering
    \includegraphics[width=13cm]{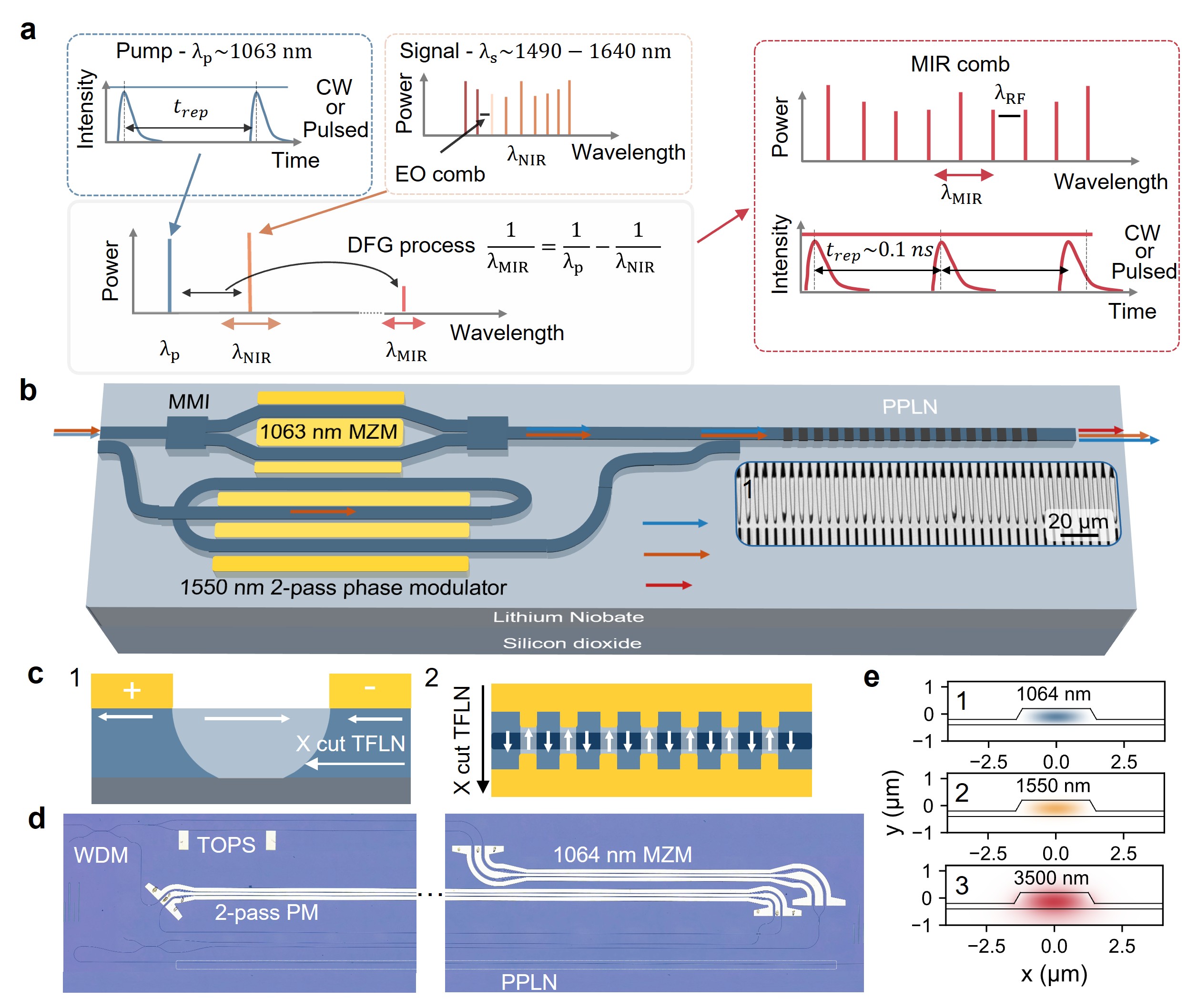}
    \caption{\textbf{Working principle of the demonstrated mid-infrared comb generator.}
    \textbf{a,} Conceptual diagram of the conversion process. A continuous-wave pump at \(1063~\mathrm{nm}\) and a tunable signal around \(1550~\mathrm{nm}\) are coupled into a thin-film lithium niobate chip, where a frequency comb near \(1550~\mathrm{nm}\) is generated using a double-pass phase-modulation scheme. An integrated Mach-Zehnder modulator on the pump arm can be used to apply intensity modulation, thereby generating mid-infrared pulses for lock-in detection, or to set the device at maximum transmission using a thermo-optic phase shifter (TOPS). The pump and comb are subsequently combined with a wavelength-division multiplexer (WDM) and routed to a PPLN section, where difference-frequency generation produces a mid-infrared frequency comb.
    \textbf{b,} Schematic of the fabricated chip, highlighting the functional sections used for pump modulation, comb preparation, wavelength combining, and nonlinear conversion in the PPLN section. The inset 1 shows a two-photon microscopy image of the poled region, revealing the domain inversion. The dark line separation highlights the location where the domain is inverted.
    \textbf{c,} Illustration of the poling process. Finger-shaped electrodes are deposited on the lithium niobate surface, and high voltage pulses up to a kilovolt are applied to invert the ferroelectric domains and form the periodically poled region. The poling period is defined by the electrode finger spacing. After poling, the lithium niobate is etched to form the ridge waveguide within the poled region.
    \textbf{d,} Optical microscope image of a fabricated device.
    \textbf{e,} Simulated fundamental TE modes in the PPLN waveguide (ridge width \(2.5~\mu\mathrm{m}\)) at the three wavelengths of interest, ie 1063 (in blue), 1550 (in orange) and around 3500 nm (in red).} 
    \label{fig:1}
\end{figure}

\section{Results}\label{sec_Result}

\subsection{Device presentation}

Figures~\ref{fig:1}a. and ~\ref{fig:1}b highlight the working principle of the device and the device schematic, respectively. At the input, the continuous wave (CW) pump and signal are coupled onto the chip into a 0.8~\textmu m wide waveguide, where the waveguide is transverse-electric single-mode at 1550~nm for efficient phase modulation. Next, pump and signal are split using a directional coupler. The pump is sent through an MZM, allowing operation either in continuous-wave or amplitude-modulated mode. The signal is sent through a double-pass phase modulator to generate an EO comb. A second directional coupler subsequently recombines the pump and signal. Next, a taper increases the waveguide top width to \(2.5~\mu\mathrm{m}\). This width is chosen as a compromise between minimizing MIR losses due to mode delocalization and maintaining an efficient MIR DFG. The taper length is optimized to avoid mode crossings around \(1~\mu\mathrm{m}\) (see Supplementary Material~C) and to minimize reflections. Finally, the pump and signal are sent through a PPLN region, whereby the signal EO comb is converted into the MIR by DFG. This periodic domain inversion modulates the sign of the nonlinear coefficient which can be used to engineer quasi-phase matching~\cite{miller1998periodically}. 
The PPLN region is formed by inverting the crystal domains in the nonlinear interaction region. Periodic poling is implemented on the unetched film by first depositing electrodes, followed by the application of high-voltage pulses to invert the ferroelectric domains in selected regions of lithium niobate, as illustrated in Figure~\ref{fig:1}c. Inset~1 of the figure shows a schematic side view of the poled structure, where the grey region indicates the inverted domain, while Inset~2 presents a top view of the inverted regions, with the white arrows indicating the orientation of the ferroelectric domain. The domain inversion is verified using two-photon microscopy, as shown in Inset~1 of Figure~\ref{fig:1}b, after which the poling electrodes are removed. 
The poling period in the PPLN region, around 4.9~\textmu m, is chosen as a compromise between nonlinear conversion efficiency, optical confinement of the pump and signal, and mitigation of MIR propagation losses. The corresponding optical modes in the PPLN region are shown in Figure~\ref{fig:1}e, and additional simulation results are provided in Supplementary Material~C. Two designs are considered: a segmented structure with four different poling periods to achieve broadband phase matching around 3.2~\textmu m (v1), and a single-period section providing phase matching centered at 3.6~\textmu m (v2). A complete optical micrograph of the chip is provided in Figure~\ref{fig:1}d.

\begin{figure}[ht!]
    \centering
    \includegraphics[width=13cm]{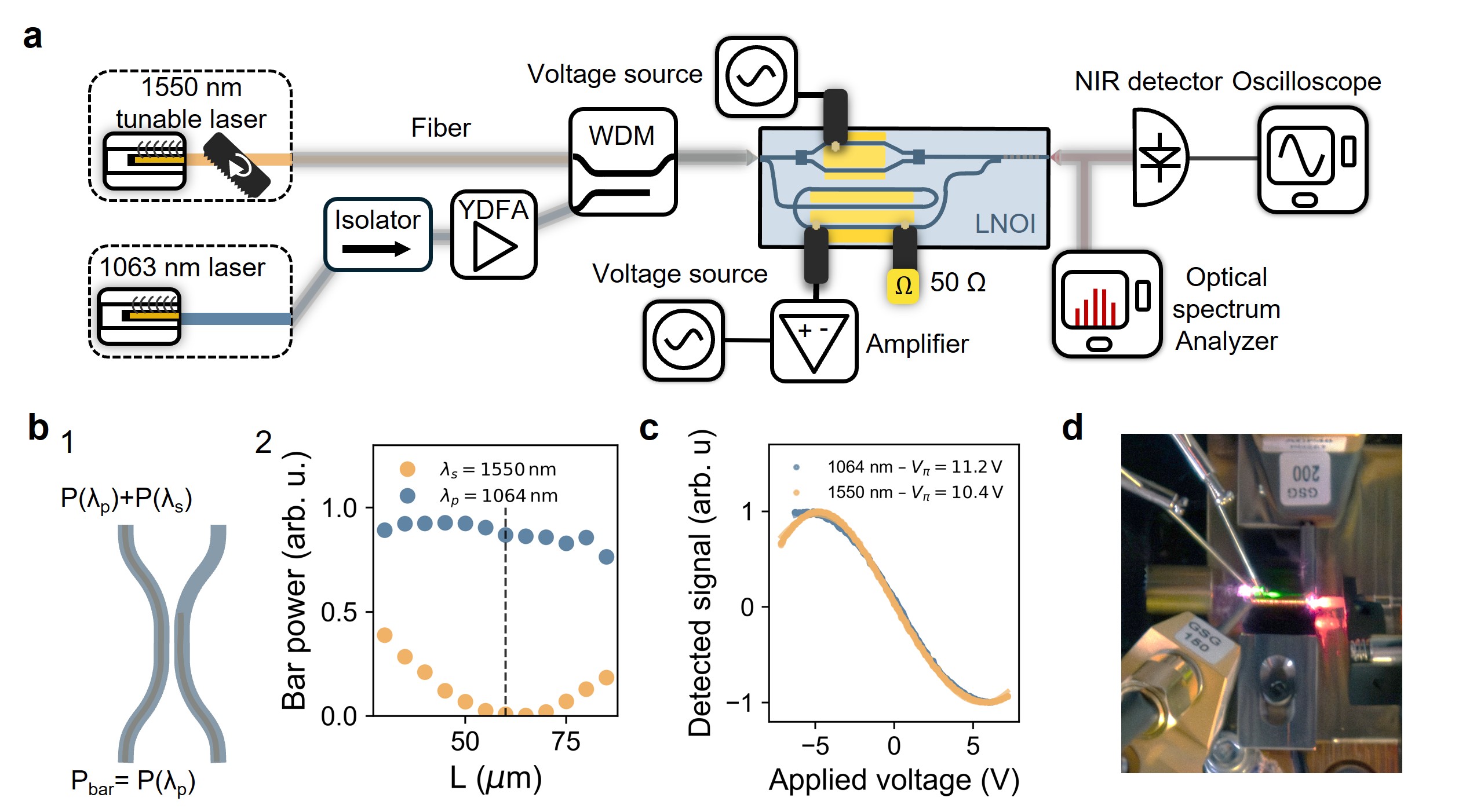}
    \caption{\textbf{Characterization setup for the MIR integrated comb generator.} 
    \textbf{a,} Schematic of the setup used to characterize the fabricated device. The pump source delivers up to 300~mW at 1063~nm, while a tunable laser source provides the signal near 1550~nm; additional details on the sources are given in the Methods section. The two signals are combined using a fiber-based wavelength-division multiplexer and injected through a polarization-maintaining lensed fiber aligned to excite the transverse-electric (TE) mode of the chip. At the output, the light is collected using a second indium-fluoride lensed fiber. The collected signal is either sent directly to an optical spectrum analyzer (OSA) or a power meter, or alternatively collimated and focused onto a high-speed NIR or MIR photodetector. Three ground-signal-ground (GSG) probes and two single-ended probes are used to drive the two-pass phase modulator, provide a 50~\(\Omega\) termination, modulate the pump when required, and bias the TOPS. 
    \textbf{b,} Directional coupler characterization for the multiplexing and de-multiplexing of the pump and signal as depicted in the panel 1. The transmitted bar power at 1063~nm is approximately 90\%, while the signal near 1550~nm is almost entirely coupled into the adjacent waveguide. 
    \textbf{c,} Voltage-length product characterization of the 1063~nm and 1550~nm modulators using an MZM test device with a length of 0.4~cm. The \(V_{\pi}L\) values of both devices are similar, at approximately 4~V$\cdot$cm. 
    \textbf{d,} Photograph of the experimental setup in operation. A two-GSG-probe configuration is used to drive and terminate the integrated microwave electrode, while two additional probes are used to set the operating point of the 1063~nm MZM. On the left side of the photograph, the input lensed fiber is visible, while the mid-infrared output lensed fiber is located on the right. The generated green light originates from second-harmonic generation of the 1063~nm pump, whereas the red emission arises from sum-frequency generation between the 1063~nm and 1550~nm signals.}
    \label{fig:2}
\end{figure}

\subsection{Static device performance characterization}

Figure~\ref{fig:2}a shows the experimental setup used to characterize the full device. Light is coupled into the fundamental TE mode of the waveguide using polarization-maintaining lensed fibers. At the output, light is collected with an InF\textsubscript{3} mid-infrared lensed fiber which is routed to an OSA or a power meter, or collimated and focused onto a mid-infrared photodetector using a fiber collimator and a high numerical-aperture germanium lens.
An important component of the platform is the on-chip 1063/1550\,nm DC wavelength division multiplexer (WDM). As shown in Figure~\ref{fig:2}b, its performance was characterized using dedicated test structures and grating couplers. By design, the 1063\,nm light remains predominantly in the input (bar) waveguide with an efficiency of approximately 80\%, whereas the 1550\,nm light is almost fully coupled to the cross waveguide, allowing efficient splitting of the two wavelengths.
The modulator exhibits a \(V_{\pi}L\) of \(V_{\pi}L = 4.2~\text{V}\,\text{cm}\) and 4.4\,V\,cm at 1550\,nm and 1063\,nm, respectively, for an electrode length of 0.4\,cm and an electrode gap of 5.8\,\(\mu\)m. The slightly lower \(V_{\pi}\) at 1063\,nm is attributed to the waveguide not being fully single mode at this wavelength. The effective \(V_{\pi}L\) of the two-pass phase modulator is \(V_{\pi}L = 4.2~\text{V}\,\text{cm}\), since it is operated in a single-ended rather than push-pull configuration. The two-pass modulator is deliberately kept simple compared to other non-resonant comb architectures~\cite{zhang2019broadband}. This is because the comb bandwidth is not the primary design target as the comb center frequency can be tuned with the input signal wavelength. The fiber-to-fiber insertion losses of the device are estimated to be approximately 8\,dB at 1063\,nm and 15\,dB at 1550\,nm. The higher loss at 1550\,nm is attributed to plasmonic losses from the long phase shifter required for comb generation. Additional platform characterization is provided in the supplementary material B.

\subsection{MIR electro-optic comb generation and nonlinear conversion.}

\begin{figure}[ht!]
    \centering
    \includegraphics[width=13.5cm]{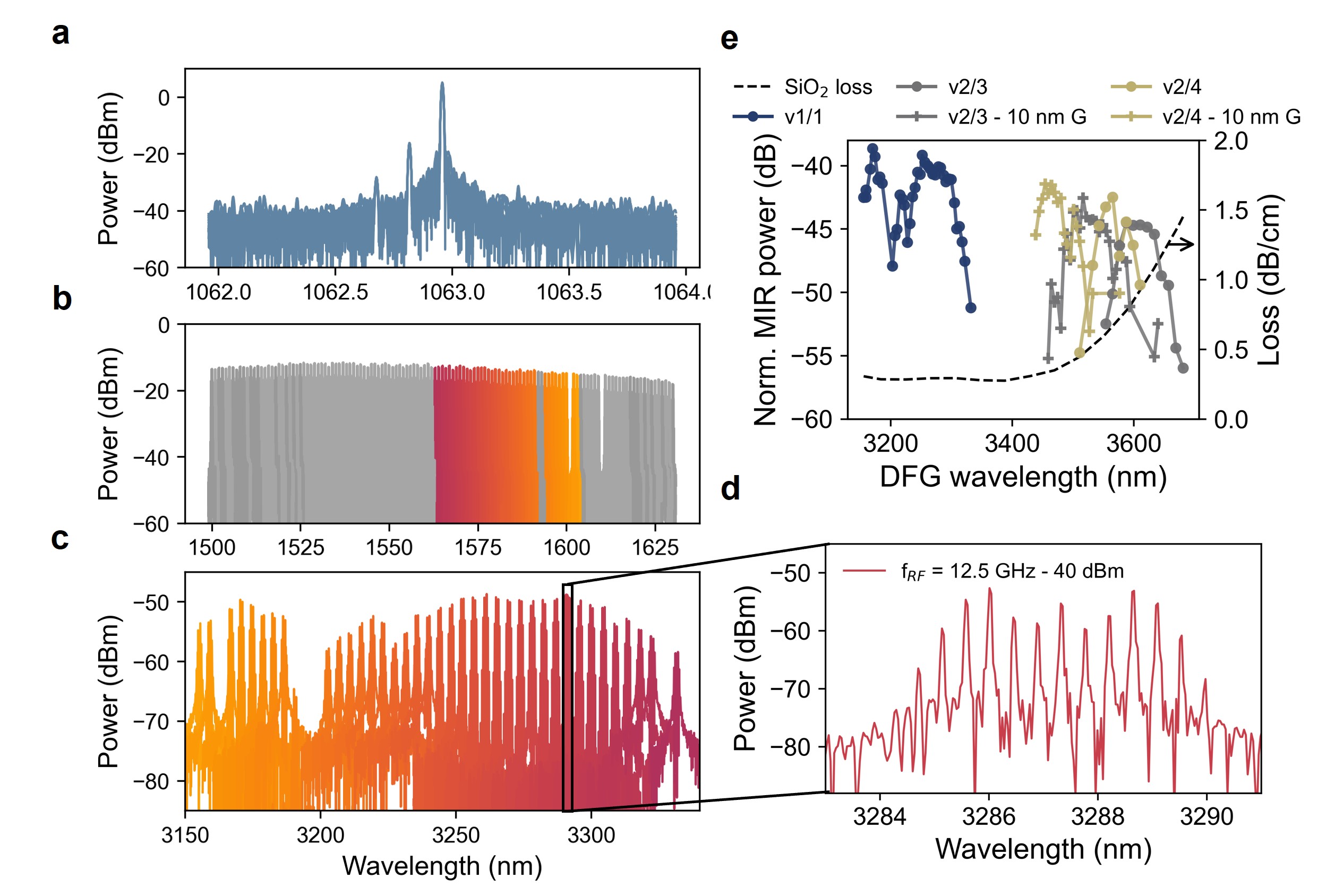}
    \caption{\textbf{Experimental demonstration of MIR comb generation with broadband difference-frequency generation.}
    \textbf{a,} Recorded output spectrum showing the pump, which remains fixed at approximately 1063~nm (purple trace), while laser-induced side modes appear around the central line. In this configuration the TOPS is set to achieve maximum transmission for the MZM.
    \textbf{b,} Recorded spectra of the modulated 1550~nm signal for different near-infrared wavelengths, acquired in 1~nm steps using a tunable 1550~nm external-cavity laser. The wavelength is swept from 1490 to 1640~nm with an output power of approximately 10~mW. In contrast to the grey curve, the colored curves from red to orange indicate the signal wavelengths for which mid-infrared light is generated, corresponding to phase-matched conditions in the PPLN.
    \textbf{c,} Corresponding MIR output spectra. The MIR emission is relatively flat from 3150 to 3300~nm, with a dip that likely arises from a region where phase matching is not satisfied.
    \textbf{d,} Example of an electro-optic frequency comb generated in the MIR with 40~dBm RF drive power, centered around 3.287~\textmu m at a modulation frequency of 12.5~GHz.
    \textbf{e,} Illustration of the MIR wavelength coverage provided by the two device platforms (v1 and v2), spanning approximately 3.0-3.69~\textmu m. The graph shows the MIR power normalized to the transmitted pump and signal at the chip output. Version~1 incorporates four distinct poling periods, while version~2 employs a single, longer poling region with multiple poling periods to extend the accessible wavelength range. Additionally, the letter G denotes devices in which a 10-nm-thick SiO$_2$ layer is deposited on top of the PPLN section. }
    \label{fig:3}
\end{figure}

Output spectra were recorded using an OSA for a device with the TOPS biased for maximum pump transmission and driven by a 12.5~GHz RF modulation signal at 40~dBm, as shown in Figure~\ref{fig:3}. The pump spectrum, shown in Figure~\ref{fig:3}a, exhibits single-mode operation around 1063~nm. Additional peaks are attributed to sidebands generated by the laser diode that were not fully suppressed by the distributed Bragg reflector structure. The signal wavelength sweep, presented in Figure~\ref{fig:3}b, demonstrates comb generation with a center wavelength tuned from 1500 to 1630~nm. The colored traces mark the wavelengths at which the phase-matching condition is satisfied and MIR emission is observed. 
Figure~\ref{fig:3}c shows the generation of an MIR DFG comb in a device consisting of four segments with different poling periods over a total length of 0.2~cm. Assuming uncorrelated pump and signal lasers, each with a linewidth of approximately 5~MHz, the linewidth of the generated MIR signal is expected to be on the order of 10~MHz, with additional possible noise coming from the elecro-optic process. This segmented quasi-phase-matching scheme enables broadband DFG, yielding MIR comb generation across a spectral bandwidth of approximately 200~nm around a central wavelength 3.2~\textmu m, at a maximum peak power of around -50~dBm. A dip in the MIR spectrum is attributed to a quasi-phase-matching mismatch between two of the four poled sections due to fabrication uncertainty of the poling period. A representative spectrum of a single comb, shown in Figure~\ref{fig:3}d, highlights an electro-optic frequency comb centered at 3.287~\textmu m, consisting of 12 sidebands spaced by 12.5~GHz and spanning a total bandwidth of approximately 6~nm with power per line around -52 dBm. In addition, higher-order modes from the signal and the pump that may arise in the taper section do not contribute to MIR generation as they are not phase matched.
Moreover, for comparison, additional PPLN designs were also investigated (design v2). The generation of MIR of around 3.6~\textmu m was achieved using a single pole section with a length of 0.8 cm as shown in Figure~\ref{fig:3}e. The MIR power was normalized to the total input power at 1063~nm and 1550~nm, as described in the Methods section. The increased interaction length and fixed poling period leads to a narrower phase-matching bandwidth and therefore to a more spectrally confined comb centered at 3.6~\textmu m, with a comb generation bandwidth of approximately 100~nm. However, optical losses in silicon dioxide increase significantly past 3.4~\textmu m, which likely accounts for the reduced conversion efficiency and lower comb power despite the longer PPLN section and the intrinsically higher gain expected from the increased interaction length. Since the conversion efficiency decreases at longer phase-matched wavelengths, the phase matching was shifted towards shorter MIR wavelengths by depositing a 10~nm glass cladding. Measurements with these samples are denoted with the letter G. Temperature tuning of the chip can also be used to achieve efficient phase matching over a broader bandwidth. The phase-matched idler wavelength is tunable by \(0.52\,\mathrm{nm/^\circ C}\), and, within the maximum operating temperature of \(160\,^\circ\mathrm{C}\) imposed by our setup, we can red-shift the MIR emission by up to \(70\,\mathrm{nm}\). More detailed measurements are provided in Supplementary material~C.

\begin{figure}[ht!]
    \centering
    \includegraphics[width=13cm]{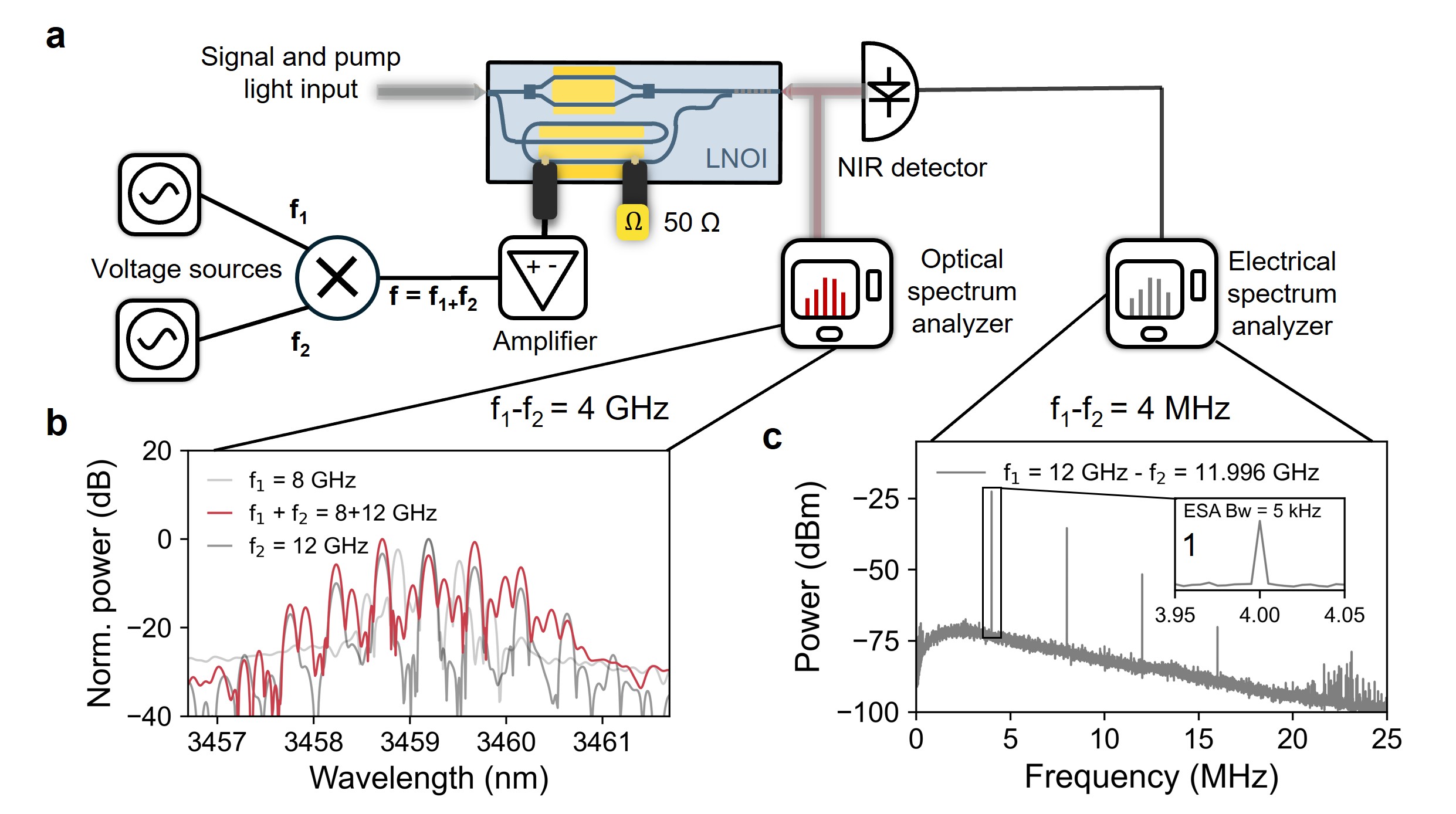}
    \caption{\textbf{Experimental demonstration of dual-tone comb generation.}
    \textbf{a,} Experimental setup used for dual-tone comb generation. Two phase-locked voltage sources are combined using an RF combiner and subsequently amplified to drive a double-pass phase modulator via a GSG probe.
    \textbf{b,} Optical spectrum analyzer measurement showing the first- and second-order sidebands corresponding to electro-optic modulation at 8~GHz and 12~GHz. The modulation frequencies are chosen such that the comb lines are resolvable with the 1.9~GHz resolution of the OSA.  
    \textbf{c,} Electrical spectrum analyzer measurement of the beat note between two signals spaced by 4~MHz, corresponding to modulation frequencies of 12~GHz and 11.996~GHz.}
    \label{fig:4}
\end{figure}

\subsection{MIR Dual-tone EO comb generation}

Interestingly, our design supports the generation of dual-tone combs with two distinct FSRs by driving the phase modulator with a two slightly different RF frequency. In this case the signal has been amplified up to 25 dBm output power to obtain more MIR DFG power. By resolving the heterodyne beat notes on the detector, the absorption spectrum can be reconstructed. As shown in Figure~\ref{fig:4}a, two phase-locked RF voltage sources were used. Their outputs were combined using an RF splitter/combiner and subsequently amplified with a single RF amplifier. The resulting RF signal was applied to the phase modulator, and the generated MIR dual-comb spectrum was observed on the optical spectrum analyzer. As seen in Figure~\ref{fig:4}b, we demonstrate the generation of a dual-tone comb with sideband spacings of 8~GHz and 12~GHz. These spacings were chosen to be well resolved by the optical spectrum analyzer, which has a resolution of approximately 1.9~GHz.
To confirm the presence of the dual-tone comb state, we measured the corresponding beat notes on a NIR photodiode. The difference between the two comb spacings was chosen to be approximately 4~MHz, so as to remain within the limited photodiode bandwidth of about 20~MHz. As highlighted in Figure~\ref{fig:4}c, the beat notes are clearly resolved. A sufficiently sensitive low-bandwidth MIR photodiode was not available for these measurements.

\section{Discussion}\label{sec12}

Our broadband integrated electro-optic comb generator provides a highly flexible route to mid-infrared comb generation, enabling independent tuning of the center wavelength over about 200\,nm (that can be extended by thermal tuning) and of the free spectral range up to 12.5\,GHz within a single device. The comb can also be operated in either continuous-wave or pulsed regimes, depending on whether the pump MZM is RF-driven, paving the way towards lock-in detection schemes. Unlike previous MIR comb demonstrations, our architecture enables simultaneous and independent control of the comb spacing (set by the RF modulation frequency) and the carrier frequency (set by the telecom pump wavelength) in a compact integrated lithium niobate platform. In contrast to earlier approaches that typically relied on high-power femtosecond pump sources, our scheme operates with two comparatively lightweight continuous-wave diode lasers delivering 24\,dBm at 1063\,nm and 10\,dBm at 1550\,nm, benefiting from the low power consumption of mature source technologies at NIR and telecom wavelengths. The MIR comb also benefits from the high spectral purity given by the intrinsic advantages of EO combs. Indeed, electro-optic combs inherit coherence only from the seed laser and RF drive, which are already well-established technologies, although coherence can still be degraded by phase noise, timing jitter, and thermal drift. Finally, the high modulation efficiency at telecom wavelengths enables dual‑tone RF driving for future dual‑comb operation using two distinct phase modulation tones, allowing the optical spectrum to be down‑converted to low‑frequency beat notes that can be detected with a low‑bandwidth MIR photodetector.

This architecture offers both wide spectral accessibility and precise scanning of molecular absorption features through RF control of the comb spacing. This is particularly advantageous for spectroscopy, where molecular absorption features often occupy only a narrow spectral window, because the comb center frequency can be tuned to concentrate the available power per comb tooth within the spectral region of interest. The possibility to easily generate dual-comb spectroscopy easily is attractive for spectroscopic applications, as it eliminates the need for a bulky high-resolution optical spectrum analyzer and instead requires only a low-bandwidth photodiode together with a modest-bandwidth electrical spectrum analyzer (ESA). The platform is further attractive for multimodal sensing because the NIR and MIR combs are simultaneously available, enabling complementary detection of molecules that exhibit absorption features in both spectral regions. Notable examples include methane, with commonly used bands near 1.65 and 3.3 \textmu m, as well as acetylene and HCN, which also show absorption features in the 1.5-\textmu m region and around 3 \textmu m. Looking forward, lithium-niobate-on-sapphire devices with thicker films could extend the accessible MIR range to about 4.4 \textmu m and improve efficiency. To improve efficiency further, adaptive poling offers a promising route to enhance the DFG efficiency and increase the generated idler power, albeit at the cost of reduced bandwidth. These results establish a scalable path towards broadband, tunable, and practically deployable mid-infrared comb sources on chip.

\section{Methods}\label{sec_Method}

\subsection{Device fabrication}

The devices are fabricated from commercially available x-cut LNOI wafers (NanoLN) comprising a 0.6~\textmu m lithium niobate thin film with a 2~\textmu m buried oxide layer. Periodic poling is first performed on the unetched film by defining chromium poling electrodes via electron-beam lithography and metal lift-off with the desired poling period. High-voltage pulses are then applied to invert the ferroelectric domains in selected regions of the lithium niobate. The domain inversion is verified using two-photon microscopy, after which the poling electrodes are removed.
Photonic waveguides are subsequently patterned in hydrogen silsesquioxane (HSQ) resist by electron-beam lithography and etched with argon etching in an ICP-RIE, yielding waveguides with top widths of 0.8~\textmu m and 2.5~\textmu m in the light-preparation and frequency-conversion sections, respectively, and an etch depth of 400~nm. A KOH cleaning step is performed to remove etch redeposition, followed by a buffered oxide etch to strip any residual mask. The anisotropic physical etch produces sidewall angles of approximately \(60^\circ\). To reduce etch-induced damage and propagation losses, the chip is then annealed at \(500\,^{\circ}\mathrm{C}\) under ambient conditions for 2~hours.
Thermo-optic phase shifters are defined by electron-beam lithography using a PMMA/MMA bilayer, followed by a 100~nm gold evaporation (Evatec~50) and lift-off. The electro-optic electrodes are patterned in a final step by direct laser writing (DWL~66+), followed by a 900~nm gold evaporation and lift-off, and are designed to provide a 50~\(\Omega\) microwave impedance to support high-speed modulation. Finally, the chip is diced (Disco DAD~3221) to expose the waveguide facets, which are mechanically polished (Allied Multiprep System 8\") using diamond abrasive to minimize in- and out-coupling losses, and further refined by focused-ion-beam milling (TFS Helios 5 UX) to improve facet quality.

\subsection{Design of the device}

The passive photonic components, namely the waveguides, MMIs, crossing and grating coupler for characterization, were first analyzed with an eigenmode expansion and finite element solver (Ansys Lumerical) to determine suitable geometries. Electro-optic performance was then evaluated with COMSOL Multiphysics and Ansys Lumerical, focusing on modulation efficiency, RF–optical field overlap, and thermo-optic behavior. Starting from the waveguide geometry optimized for single-mode guidance and a chosen gap, the widths and spacings of the transmission lines were refined through RF–optical co-simulations in COMSOL Multiphysics to achieve 50~$\Omega$ impedance and velocity matching with the optical mode, thereby suppressing reflections and enhancing the modulation bandwidth. Previous characterization has confirmed that the presented radio-frequency design demonstrates bandwidths exceeding 20 GHz.


\subsection{Electro-optic characterization}

The pump source is a distributed Bragg reflector (DBR) laser emitting near 1063~nm, with a maximum output power of 10~mW (Thorlabs DBR1064P) and a linewidth of approximately 5~MHz. Its output is subsequently amplified by a polarization-maintaining YDFA with a saturation power of 26~dBm (Civil Laser YDFA-26-PM-M). The signal source is a TOPTICA DLC laser delivering up to 30~mW over the 1460--1560~nm wavelength range. Although its observed linewidth is approximately 5~MHz due to frequency drift, its intrinsic linewidth is on the order of 100~Hz. In specific cases, a Keysight Newport laser (N7776C) was used to access longer wavelengths from 1490 to 1640~nm, providing an output power of around 10~mW and a linewidth of approximately 50~MHz. Optionally, the signal can be amplified using an EDFA (CEFA-C-HG) with a saturation power of 33~dBm but only power up to 25 dBm have been used.
To evaluate the modulation efficiency, static measurements were performed on a standalone test MZM using grating-coupler coupling. The optical path difference between the two arms is thermally tunable via the integrated TOPS, which was characterized by sweeping the electrical power and monitoring the resulting change in the MZM output power. The half-wave voltages \(V_{\pi}\) in \(1063\,\mathrm{nm}\) and \(1550\,\mathrm{nm}\) were extracted by driving the modulator with a triangular waveform \(1\,\mathrm{kHz}\) from an arbitrary function generator, amplified to \(200\,\mathrm{V}\) using a high-voltage amplifier. The modulated optical signal was detected with an InGaAs photodetector (New Focus 2053), recorded on a digital oscilloscope, and analyzed using a custom Python routine, as shown in Fig.~\ref{fig:2}c.

\subsection{MIR electro-optic comb generation and characterization}

For efficient signal comb generation, we implemented a double-pass phase modulation scheme, as illustrated in Fig.~\ref{fig:1}b, which effectively increases the modulation depth without requiring higher RF drive power or a longer device. Radio-frequency signals were generated externally, amplified using a high-power RF amplifier (RF-Lambda RFLUPA06G12GB), and delivered to the on-chip traveling-wave electrodes via a high-speed GSG probe. The electrode termination was matched to \(50~\Omega\) to suppress RF reflections.
The electrode spacing was carefully optimized to minimize plasmonic losses while maintaining efficient electro-optic interaction, which is critical for preserving optical power at 1550~nm for subsequent DFG. In addition, the electrode width and spacing were optimized to achieve \(50~\Omega\) impedance matching and to match the effective indices of the optical and electrical modes. Since the comb bandwidth scales with the phase modulation index, the double-pass configuration increases the number of generated sidebands for a given RF drive power.
Device design and optimization were performed using COMSOL Multiphysics and Ansys Lumerical, enabling evaluation of modulation efficiency, optical mode confinement, and thermo-optic effects.
To obtain the spectra of the combs at different center wavelengths, we swept the signal wavelength and recorded the corresponding output spectra using an OSA. The modulated output was collected using a second lensed fiber and routed to the OSA for spectral analysis. To compare measurements acquired at different pump and signal powers, which may vary due to facet quality and device-to-device variations, we normalized the measured MIR power to the total input power at 1063~nm and at the considered NIR wavelength \(\lambda_{\mathrm{NIR}}\). The normalized MIR power is defined as
\begin{equation}
P_{\mathrm{MIR,norm}}(\lambda_{\mathrm{MIR}})
= P_{\mathrm{MIR}}(\lambda_{\mathrm{MIR}}) - P_{\mathrm{in}},
\end{equation}
where
\begin{equation}
P_{\mathrm{in}}
= 10 \log_{10}\!\left(10^{P_{1063}/10} + 10^{P_{\mathrm{NIR}}(\lambda_{\mathrm{NIR}})/10}\right).
\end{equation}
Here, \(P_{\mathrm{MIR}}(\lambda_{\mathrm{MIR}})\) is the measured MIR power at the DFG wavelength \(\lambda_{\mathrm{MIR}}\) corresponding to the NIR wavelength \(\lambda_{\mathrm{NIR}}\), while \(P_{1063}\) and \(P_{\mathrm{NIR}}(\lambda_{\mathrm{NIR}})\) are the measured input powers in dBm at 1063~nm and at \(\lambda_{\mathrm{NIR}}\), respectively. Consequently, \(P_{\mathrm{MIR,norm}}(\lambda_{\mathrm{MIR}})\) is expressed in dB.


\backmatter

\bmhead{Supplementary information}

Further experimental data and analysis can be found in the Supplementary Information.

\bmhead{Acknowledgements}

The authors appreciate support for the fabrication and characterization of our samples from the Scientific Center for Optical and Electron Microscopy (ScopeM), the BRNC and FIRST cleanroom facilities at ETH Zurich, and IBM Rüschlikon. We acknowledge support from the Schweizerischer Nationalfonds (SNF) through a postdoctoral fellowship under the HARMONY project and Ambizione Fellowship 208707. We acknowledge funding for a PhD position from ETH Zürich within the project 23-2 ETH-33 - MIPS4LIFE (Mid-Infrared Photonics For the LIFE Space Mission). We thank Dr. Antonis Olziersky for support with the electron beam lithography process. Finally, we thank Dr. Mathieu Bertrand for fruitful scientific discussions.

\bmhead{Data availability}

The data supporting the plots within this article are available from the corresponding author upon request.


\noindent





\begin{appendices}

\section{Platform fabrication process}\label{app4}

\begin{figure}[H]
    \centering
    \includegraphics[width=13cm]{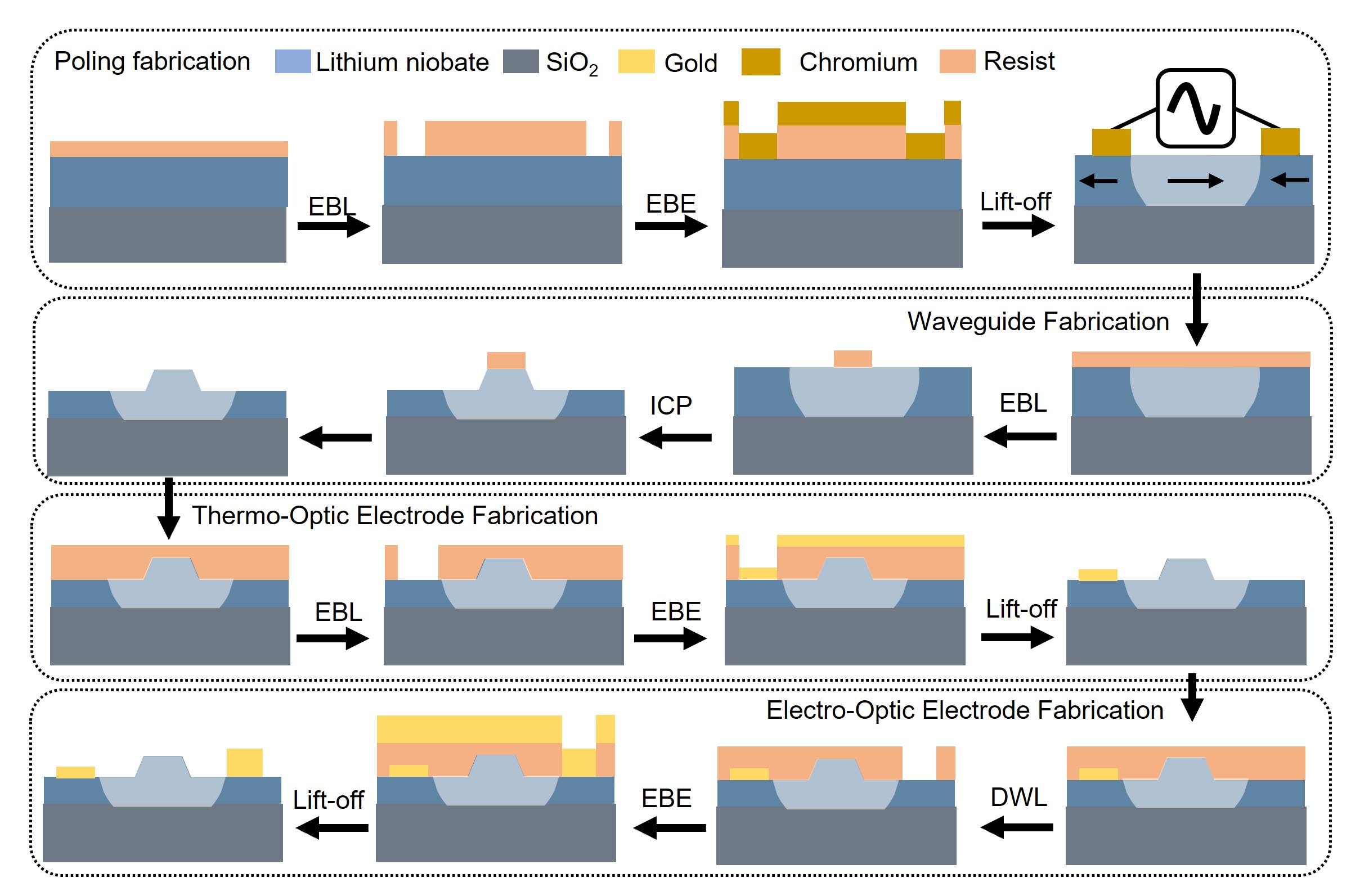}
    \caption{\textbf{Fabrication process of the TFLN on insulator} 
    Fabrication process for lithium niobate on insulator with PPLN section.
    EBL: electron-beam lithography; ICP: inductively coupled plasma; EBE: electron-beam evaporation; DWL: direct-write laser lithography.}
    \label{fig:1_supp}
\end{figure}

Supplementary Figure~\ref{fig:1_supp} presents a schematic of the fabrication flow, described in detail in the Methods section, used to realize thin-film lithium niobate on insulator photonic circuits incorporating a PPLN section.

\section{Additional platform characterization}\label{app4}

For component characterization, light at 1063~nm and 1550~nm is coupled on and off the chip using optimized linear grating couplers, ensuring stable coupling conditions and improved measurement reproducibility. The MMI performance is evaluated with four-stage cascaded MMI structures terminated by grating couplers at 1063~nm; the corresponding results are shown in Supplementary Figure~\ref{fig:2_supp}a. From a linear fit of the transmitted power, the excess loss of the MMIs is extracted to be 0.7~dB at 1063~nm. The grating couplers are first designed using a commercial FDTD solver, where parameter sweeps over grating period, duty cycle, and emission angle are performed to identify optimal configurations for the target wavelength and polarization. The measured grating-coupler response around 1550~nm is presented in Supplementary Figure~\ref{fig:2_supp}b, showing minimum coupling losses of approximately 4~dB per grating at 1550~nm. The crossing loss is characterized using four-stage cascaded crossings placed after a 1×2 MMI splitter, which allows us to remove the contribution of the grating couplers and use a reference arm with identical couplers at 1550~nm; the results in Supplementary Figure~\ref{fig:2_supp}c indicate that the crossing loss remains below 2~dB over the full wavelength range, and below 1~dB between 1500 and 1600~nm. Finally, the bar-port power at 1550~nm is measured for a 2×2 directional coupler, as reported in Supplementary Figure~\ref{fig:2_supp}d; the very low bar-port transmission confirms the good performance of the wavelength-division multiplexing scheme with 1063~nm and 1550~nm signals.

\begin{figure}[H]
    \centering
    \includegraphics[width=13cm]{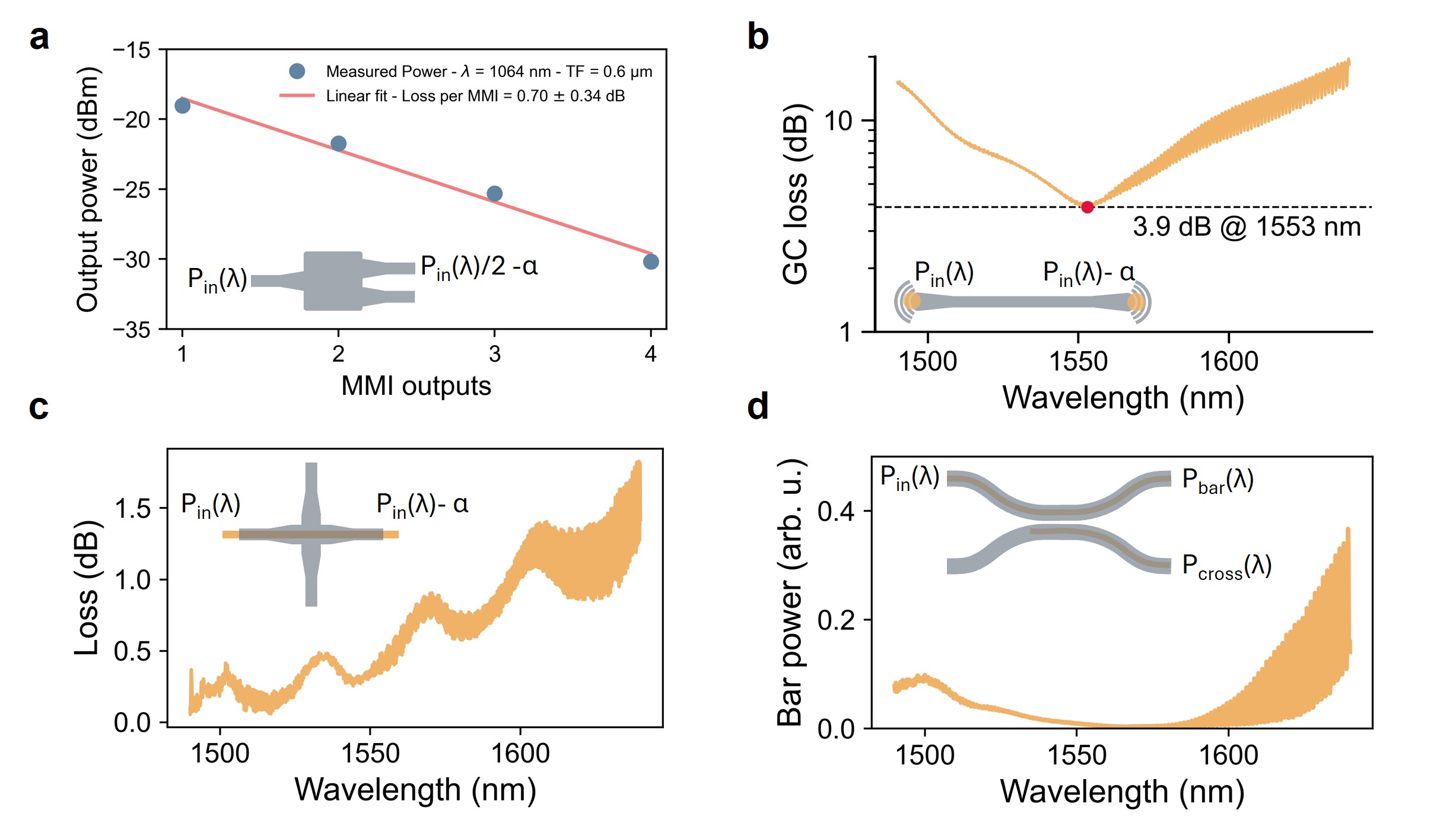}
    \caption{
    \textbf{Characterization of passive components.}
    \textbf{a,} Transmission of four-stage cascaded MMIs at 1063~nm, used to extract an excess MMI loss of 0.7~dB. 
    \textbf{b,} Measured grating-coupler response around 1550~nm, showing a minimum coupling loss of about 4~dB per grating. 
    \textbf{c,} Crossing-loss measurement using four-stage cascaded crossings with a 1$\times$2 MMI reference arm, demonstrating losses below 2~dB over the full wavelength range and below 1~dB between 1500 and 1600~nm. 
    \textbf{d,} Bar-port power at 1550~nm for a 2$\times$2 directional coupler, indicating negligible bar-port transmission and efficient wavelength-division multiplexing between 1063~nm and 1550~nm.
    }
    \label{fig:2_supp}
\end{figure}

\section{Nonlinear PPLN simulation}\label{app4}

\begin{figure}[H]
    \centering
    \includegraphics[width=13cm]{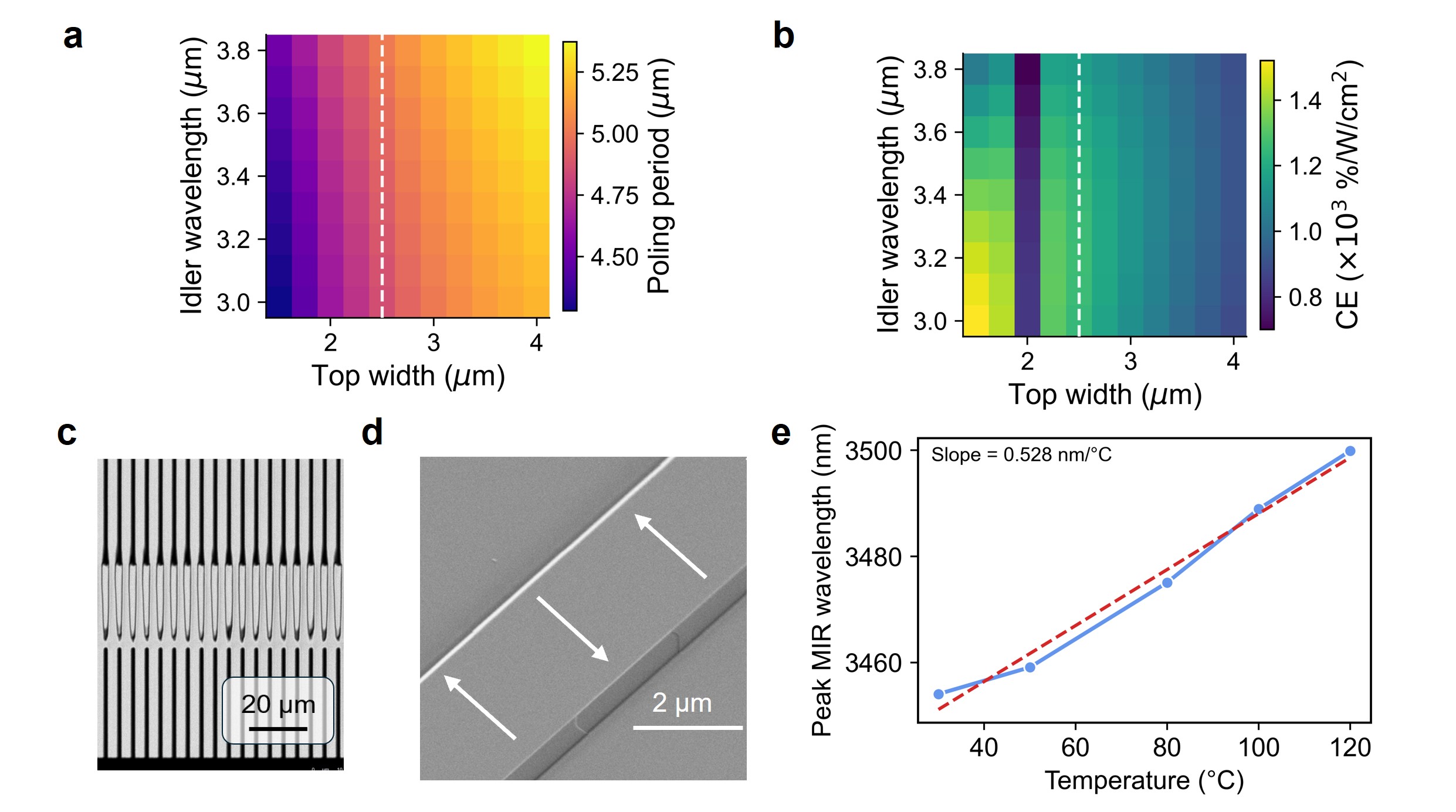}
    \caption{\textbf{Simulated dependence of the PPLN waveguide design on top width and idler wavelength. }
    \textbf{a,} Required poling period to achieve quasi-phase-matching as a function of idler wavelength and waveguide top width. 
    \textbf{b,} Corresponding normalized conversion efficiency (CE), highlighting a mode-crossing region with a pronounced efficiency drop. 
    To avoid weak confinement and high loss of the mid-infrared mode below this region, a waveguide width above the mode crossing is selected, with an additional safety margin to account for fabrication and modeling variations.
    \textbf{c,} Two-photon microscopy image of the poled region, revealing the domain inversion. The dark line separation highlights the location where the domain is inverted.
    \textbf{d,} SEM image of the poled region, revealing the domain inversion through the corrugation that arises because the waveguide is etched differently depending on the ferroelectric orientation. White arrows indicate the domain orientation.
    \textbf{e,} Experimental result showing the resulting DFG MIR emission peak wavelength as a function of temperature.
    }
    \label{fig:3_supp}
\end{figure}

Figure~\ref{fig:3_supp} presents additional information and characterization for the PPLN region. Figure~\ref{fig:3_supp}a shows the poling period required to satisfy the quasi-phase-matching condition as a function of idler wavelength and waveguide top width. Figure~\ref{fig:3_supp}b presents the corresponding normalized conversion efficiency (CE), revealing a mode-crossing region where the efficiency drops sharply. Below this mode-crossing region, the mid-infrared mode would be weakly confined and highly lossy; therefore, we chose a waveguide width above the mode crossing and maintained a safety margin to account for potential deviations from the simulations. In Figure~\ref{fig:3_supp}c, a two-photon microscopy image of the poled region reveals the domain inversion. The dark line separation highlights the location where the domain is inverted. Figure~\ref{fig:3_supp}d shows an SEM image of the poled region, revealing the domain inversion through corrugation that arises because the waveguide is etched differently depending on the ferroelectric orientation. White arrows indicate the domain orientation. As shown in Figure~\ref{fig:3_supp}e, changing the temperature slightly shifts the refractive indices of the lithium niobate crystal, which changes which wavelength satisfies the quasi-phase-matching condition. This allows the output MIR wavelength to be tuned by heating or cooling the crystal; higher temperature shifts the phase-matching wavelength, enabling precise control of the generated light without changing the poling period.

\end{appendices}


\bibliography{sn-bibliography}

@article{abajyan2025mid,
  title={Mid-infrared frequency combs and pulse generation based on single section interband cascade lasers},
  author={Abajyan, Pavel and Chomet, Baptiste and Diaz-Thomas, Daniel A and Saemian, Mohammadreza and Mi{\v{c}}ica, Martin and Mangeney, Juliette and Tignon, Jerome and Baranov, Alexei N and Pantzas, Konstantinos and Sagnes, Isabelle and others},
  journal={Applied Physics Letters},
  volume={126},
  number={13},
  year={2025},
  publisher={AIP Publishing}
}

@book{miller1998periodically,
  title={Periodically poled lithium niobate: modeling, fabrication, and nonlinear optical performance},
  author={Miller, Gregory David},
  year={1998},
  publisher={Stanford university}
}

@article{lee2026suspended,
  title={Suspended thin-film lithium niobate modulator for broadband mid-infrared light modulation and frequency comb generation},
  author={Lee, Chun-Ho and Ren, Xinyi and Su, Xinzhou and Lee, Wonho and Jiang, Zile and Yu, Yue and Zhou, Huibin and Zuo, Yue and Ou, Shaoyuan and Kopparapu, Reshma and others},
  journal={arXiv preprint arXiv:2601.17385},
  year={2026}
}

@article{hoghooghi2022broadband,
  title={Broadband 1-GHz mid-infrared frequency comb},
  author={Hoghooghi, Nazanin and Xing, Sida and Chang, Peter and Lesko, Daniel and Lind, Alexander and Rieker, Greg and Diddams, Scott},
  journal={Light: Science \& Applications},
  volume={11},
  number={1},
  pages={264},
  year={2022},
  publisher={Nature Publishing Group UK London}
}

@article{roy2023visible,
  title={Visible-to-mid-IR tunable frequency comb in nanophotonics},
  author={Roy, Arkadev and Ledezma, Luis and Costa, Luis and Gray, Robert and Sekine, Ryoto and Guo, Qiushi and Liu, Mingchen and Briggs, Ryan M and Marandi, Alireza},
  journal={Nature Communications},
  volume={14},
  number={1},
  pages={6549},
  year={2023},
  publisher={Nature Publishing Group UK London}
}

@article{bagheri2018passively,
  title={Passively mode-locked interband cascade optical frequency combs},
  author={Bagheri, Mahmood and Frez, Clifford and Sterczewski, Lukasz A and Gruidin, Ivan and Fradet, Mathieu and Vurgaftman, Igor and Canedy, Chadwick L and Bewley, William W and Merritt, Charles D and Kim, Chul Soo and others},
  journal={Scientific Reports},
  volume={8},
  number={1},
  pages={3322},
  year={2018},
  publisher={Nature Publishing Group UK London}
}

@article{coddington2016dual,
  title={Dual-comb spectroscopy},
  author={Coddington, Ian and Newbury, Nathan and Swann, William},
  journal={Optica},
  volume={3},
  number={4},
  pages={414--426},
  year={2016},
  publisher={Optical Society of America}
}

@article{del2007optical,
  title={Optical frequency comb generation from a monolithic microresonator},
  author={Del’Haye, Pascal and Schliesser, Albert and Arcizet, Olivier and Wilken, Tom and Holzwarth, Ronald and Kippenberg, Tobias J},
  journal={Nature},
  volume={450},
  number={7173},
  pages={1214--1217},
  year={2007},
  publisher={Nature Publishing Group UK London}
}

@article{diddams2007molecular,
  title={Molecular fingerprinting with the resolved modes of a femtosecond laser frequency comb},
  author={Diddams, Scott A and Hollberg, Leo and Mbele, Vela},
  journal={Nature},
  volume={445},
  number={7128},
  pages={627--630},
  year={2007},
  publisher={Nature Publishing Group UK London}
}

@article{picque2019frequency,
  title={Frequency comb spectroscopy},
  author={Picqu{\'e}, Nathalie and H{\"a}nsch, Theodor W},
  journal={Nature photonics},
  volume={13},
  number={3},
  pages={146--157},
  year={2019},
  publisher={Nature Publishing Group UK London}
}

@article{udem2002optical,
  title={Optical frequency metrology},
  author={Udem, Th and Holzwarth, Ronald and H{\"a}nsch, Theodor W},
  journal={Nature},
  volume={416},
  number={6877},
  pages={233--237},
  year={2002},
  publisher={Nature Publishing Group UK London}
}

@article{didier2026thin,
  title={Thin film lithium niobate on sapphire for integrated mid-infrared modulator},
  author={Didier, Pierre and Jain, Prakhar and Bertrand, Mathieu and Kellner, Jost and Pitz, Oliver and Dai, Zhecheng and Kuttner, Tristan and Beck, Mattias and Chen, Baile and Faist, J{\'e}r{\^o}me and others},
  journal={Nature Communications},
  year={2026},
  publisher={Nature Publishing Group UK London}
}

@article{zhang2019broadband,
  title={Broadband electro-optic frequency comb generation in a lithium niobate microring resonator},
  author={Zhang, Mian and Buscaino, Brandon and Wang, Cheng and Shams-Ansari, Amirhassan and Reimer, Christian and Zhu, Rongrong and Kahn, Joseph M and Lon{\v{c}}ar, Marko},
  journal={Nature},
  volume={568},
  number={7752},
  pages={373--377},
  year={2019},
  publisher={Nature Publishing Group UK London}
}

@article{rueda2019resonant,
  title={Resonant electro-optic frequency comb},
  author={Rueda, Alfredo and Sedlmeir, Florian and Kumari, Madhuri and Leuchs, Gerd and Schwefel, Harald GL},
  journal={Nature},
  volume={568},
  number={7752},
  pages={378--381},
  year={2019},
  publisher={Nature Publishing Group UK London}
}

@article{pfeiffer2017octave,
  title={Octave-spanning dissipative Kerr soliton frequency combs in Si3N4 microresonators},
  author={Pfeiffer, Martin HP and Herkommer, Clemens and Liu, Junqiu and Guo, Hairun and Karpov, Maxim and Lucas, Erwan and Zervas, Michael and Kippenberg, Tobias J},
  journal={Optica},
  volume={4},
  number={7},
  pages={684--691},
  year={2017},
  publisher={Optical Society of America}
}

@article{guo2018mid,
  title={Mid-infrared frequency comb via coherent dispersive wave generation in silicon nitride nanophotonic waveguides},
  author={Guo, Hairun and Herkommer, Clemens and Billat, Adrien and Grassani, Davide and Zhang, Chuankun and Pfeiffer, Martin HP and Weng, Wenle and Br{\`e}s, Camille-Sophie and Kippenberg, Tobias J},
  journal={Nature Photonics},
  volume={12},
  number={6},
  pages={330--335},
  year={2018},
  publisher={Nature Publishing Group UK London}
}

@article{wang2019monolithic,
  title={Monolithic lithium niobate photonic circuits for Kerr frequency comb generation and modulation},
  author={Wang, Cheng and Zhang, Mian and Yu, Mengjie and Zhu, Rongrong and Hu, Han and Loncar, Marko},
  journal={Nature communications},
  volume={10},
  number={1},
  pages={978},
  year={2019},
  publisher={Nature Publishing Group UK London}
}

@article{hu2022high,
  title={High-efficiency and broadband on-chip electro-optic frequency comb generators},
  author={Hu, Yaowen and Yu, Mengjie and Buscaino, Brandon and Sinclair, Neil and Zhu, Di and Cheng, Rebecca and Shams-Ansari, Amirhassan and Shao, Linbo and Zhang, Mian and Kahn, Joseph M and others},
  journal={Nature photonics},
  volume={16},
  number={10},
  pages={679--685},
  year={2022},
  publisher={Nature Publishing Group UK London}
}

@article{zhang2025ultrabroadband,
  title={Ultrabroadband integrated electro-optic frequency comb in lithium tantalate},
  author={Zhang, Junyin and Wang, Chengli and Denney, Connor and Riemensberger, Johann and Lihachev, Grigory and Hu, Jianqi and Kao, Wil and Bl{\'e}sin, Terence and Kuznetsov, Nikolai and Li, Zihan and others},
  journal={Nature},
  volume={637},
  number={8048},
  pages={1096--1103},
  year={2025},
  publisher={Nature Publishing Group UK London}
}

@article{zhang2023power,
  title={A power-efficient integrated lithium niobate electro-optic comb generator},
  author={Zhang, Ke and Sun, Wenzhao and Chen, Yikun and Feng, Hanke and Zhang, Yiwen and Chen, Zhaoxi and Wang, Cheng},
  journal={Communications Physics},
  volume={6},
  number={1},
  pages={17},
  year={2023},
  publisher={Nature Publishing Group UK London}
}

@article{chang2022integrated,
  title={Integrated optical frequency comb technologies},
  author={Chang, Lin and Liu, Songtao and Bowers, John E},
  journal={Nature Photonics},
  volume={16},
  number={2},
  pages={95--108},
  year={2022},
  publisher={Nature Publishing Group UK London}
}

@article{xie2022linbo3,
  title={LiNbO3 crystals: from bulk to film},
  author={Xie, Zhenda and Zhu, Shining},
  journal={Advanced Photonics},
  volume={4},
  number={3},
  pages={030502--030502},
  year={2022},
  publisher={Society of Photo-Optical Instrumentation Engineers}
}

@incollection{boyd2008nonlinear,
  title={Nonlinear optics},
  author={Boyd, Robert W and Gaeta, Alexander L and Giese, Enno},
  booktitle={Springer Handbook of Atomic, Molecular, and Optical Physics},
  pages={1097--1110},
  year={2008},
  publisher={Springer}
}

@article{schilt2003wavelength,
  title={Wavelength modulation spectroscopy: combined frequency and intensity laser modulation},
  author={Schilt, Stephane and Thevenaz, Luc and Robert, Philippe},
  journal={Applied optics},
  volume={42},
  number={33},
  pages={6728--6738},
  year={2003},
  publisher={Optical Society of America}
}

@article{wang2018integrated,
  title={Integrated lithium niobate electro-optic modulators operating at CMOS-compatible voltages},
  author={Wang, Cheng and Zhang, Mian and Chen, Xi and Bertrand, Maxime and Shams-Ansari, Amirhassan and Chandrasekhar, Sethumadhavan and Winzer, Peter and Lon{\v{c}}ar, Marko},
  journal={Nature},
  volume={562},
  number={7725},
  pages={101--104},
  year={2018},
  publisher={Nature Publishing Group UK London}
}

@article{xu2020high,
  title={High-performance coherent optical modulators based on thin-film lithium niobate platform},
  author={Xu, Mengyue and He, Mingbo and Zhang, Hongguang and Jian, Jian and Pan, Ying and Liu, Xiaoyue and Chen, Lifeng and Meng, Xiangyu and Chen, Hui and Li, Zhaohui and others},
  journal={Nature communications},
  volume={11},
  number={1},
  pages={3911},
  year={2020},
  publisher={Nature Publishing Group UK London}
}

@article{hugi2012mid,
  title={Mid-infrared frequency comb based on a quantum cascade laser},
  author={Hugi, Andreas and Villares, Gustavo and Blaser, St{\'e}phane and Liu, HC and Faist, J{\'e}r{\^o}me},
  journal={Nature},
  volume={492},
  number={7428},
  pages={229--233},
  year={2012},
  publisher={Nature Publishing Group UK London}
}

@article{montesinos2022mid,
  title={Mid-infrared integrated electro-optic modulator operating up to 225 MHz between 6.4 and 10.7 $\mu$m wavelength},
  author={Montesinos-Ballester, Miguel and Deniel, Lucas and Koompai, Natnicha and Nguyen, Thi Hao Nhi and Frigerio, Jacopo and Ballabio, Andrea and Falcone, Virginia and Le Roux, Xavier and Alonso-Ramos, Carlos and Vivien, Laurent and others},
  journal={ACS photonics},
  volume={9},
  number={1},
  pages={249--255},
  year={2022},
  publisher={ACS Publications}
}

@inproceedings{liu2019mid,
  title={{Mid and long-wave infrared free-space optical communication}},
  author={Liu, Jony J and Stann, Barry L and Klett, Karl K and Cho, Pak S and Pellegrino, Paul M},
  booktitle={Laser communication and propagation through the atmosphere and oceans VIII},
  volume={11133},
  pages={1113302},
  year={2019},
  organization={International Society for Optics and Photonics}
}

@article{corrigan2009quantum,
  title={{Quantum cascade lasers and the Kruse model in free space optical communication}},
  author={Corrigan, Paul and Martini, Rainer and Whittaker, Edward A and Bethea, Clyde},
  journal={Optics Express},
  volume={17},
  number={6},
  pages={4355--4359},
  year={2009},
  publisher={Optical Society of America}
}

@article{zhang2014applications,
  title={{Applications of absorption spectroscopy using quantum cascade lasers}},
  author={Zhang, Lizhu and Tian, Guang and Li, Jingsong and Yu, Benli},
  journal={Applied spectroscopy},
  volume={68},
  number={10},
  pages={1095--1107},
  year={2014},
  publisher={SAGE Publications Sage UK: London, England}
}

@article{Spitz2021,
  title={{Free-Space Communication with Directly Modulated Mid-Infrared Quantum Cascade Devices}},
  author={Spitz, Olivier and Didier, Pierre and Durupt, Laur{\'e}line and D{\'\i}az-Thomas, Daniel Andres and Baranov, Alexei N and Cerutti, Laurent and Grillot, Fr{\'e}d{\'e}ric},
  journal={IEEE Journal of Selected Topics in Quantum Electronics},
  volume={28},
  number={1},
  pages={1--9},
  year={2021},
  publisher={IEEE}
}

@article{haas2016advances,
  title={{Advances in mid-infrared spectroscopy for chemical analysis}},
  author={Haas, Julian and Mizaikoff, Boris},
  journal={Annu. Rev. Anal. Chem},
  volume={9},
  number={1},
  pages={45--68},
  year={2016}
}

@article{kazakov2025driven,
  title={Driven bright solitons on a mid-infrared laser chip},
  author={Kazakov, Dmitry and Letsou, Theodore P and Piccardo, Marco and Columbo, Lorenzo L and Brambilla, Massimo and Prati, Franco and Dal Cin, Sandro and Beiser, Maximilian and Opa{\v{c}}ak, Nikola and Ratra, Pawan and others},
  journal={Nature},
  pages={1--7},
  year={2025},
  publisher={Nature Publishing Group UK London}
}

@article{Esmail2017,
author = {Esmail, Maged Abdullah and Ragheb, Amr and Fathallah, Habib and Alouini, Mohamed Slim},
issn = {19430655},
journal = {IEEE Photonics J.},
number = {1},
title = {{Investigation and Demonstration of High Speed Full-Optical Hybrid FSO/Fiber Communication System under Light Sand Storm Condition}},
volume = {9},
year = {2017}
}

@article{Schwarz2019,
archivePrefix = {arXiv},
arxivId = {1812.03879},
author = {Schwarz, Benedikt and Hillbrand, Johannes and Beiser, Maximilian and Andrews, Aaron Maxwell and Strasser, Gottfried and Detz, Hermann and Schade, Anne and Weih, Robert and H{\"{o}}fling, Sven},
doi = {10.1364/optica.6.000890},
eprint = {1812.03879},
issn = {23342536},
journal = {Optica},
number = {7},
pages = {890},
title = {{Monolithic frequency comb platform based on interband cascade lasers and detectors}},
volume = {6},
year = {2019}
}

@article{heckelmann2023quantum,
  title={Quantum walk comb in a fast gain laser},
  author={Heckelmann, Ina and Bertrand, Mathieu and Dikopoltsev, Alexander and Beck, Mattias and Scalari, Giacomo and Faist, J{\'e}r{\^o}me},
  journal={Science},
  volume={382},
  number={6669},
  pages={434--438},
  year={2023},
  publisher={American Association for the Advancement of Science}
}

@article{Jony2019,
    author = {Jony J. Liu and Barry L. Stann and Karl K. Klett and Pak S. Cho and Paul M. Pellegrino},
    title = {{Mid and long-wave infrared free-space optical communication}},
    volume = {11133},
    journal = {Laser Communication and Propagation through the Atmosphere and Oceans VIII},
    publisher = {SPIE},
    pages = {1113302},
    year = {2019},
    URL = {https://doi.org/10.1117/12.2530969}
    }

\end{document}